\documentclass{article}
\usepackage{spconf,amsmath, amssymb,graphicx,hyperref}
\usepackage{booktabs}
\usepackage{pifont}

\title{Zephyr: An Efficient Audio Denoising System Using Spiking Neural Networks Enabled with a Sparsity-Aware Flexible FPGA PE Array}
\name{\begin{tabular}[t]{c}
Cheng-En Chang$^{\star \ddagger}$, Chi-Wei Kao$^{\star \ddagger}$ , Chung-Lun Yang$^{\star \ddagger}$, Yan-Lin Jiang$^{\star \ddagger}$, Yi-Chen Huang$^{\star \ddagger}$,\\ Sebastian Fieldhouse$^{\dagger}$ and Kea-Tiong Tang$^{\star}$
\end{tabular}
}

  \address{
  $^{\star}$ Department of Electrical Engineering, National Tsing Hua University, Taiwan \\ 
  $^{\dagger}$College of Semiconductor Research, National Tsing Hua University, Taiwan \\ 
  $^{\ddagger}$These authors contributed equally to this work.}
\begin{document}
%
\maketitle

\begin{abstract}
In this work we look to neuromorphic computing to solve the power consumption problem that audio denoising neural networks face on edge devices like smartphones, wireless headphones and hearing aids. Spiking neural networks (SNNs) have the potential to solve this problem due to their high activation sparsity and low complexity, however many SOTA SNNs require hardware that supports a mixture of operations to be able to fully perform inference. To solve this problem, we convert SOTA audio denoising neural network Spiking-FullSubNet to a hardware friendly version showing that via QAT and activation function simplification we can achieve $\approx28\times$ improvement in power consumption to $52.9$nJ per 32ms audio frame when calculated for custom digital hardware in a 45nm process node. We then propose a digital circuit which by means of a sparsity-aware flexible PE array can perform inference of the heterogeneous compute load of Spiking-FullSubNet, and validate this circuit on a PYNQ-Z1 FPGA achieving a real-time factor of 0.727 at 100MHz. 
\end{abstract}
\section{Introduction}
\label{sec:intro}
Smartphones, headphones and hearing aids are all examples of highly power constrained edge devices that are used to run audio denoising artificial neural networks (ANNs) to improve user experience. However, these devices are severely power constrained relying on battery power with users expecting long battery life between recharges, whilst ANNs are computationally resource and power hungry. Motivated by this problem we look to neuromorphic computing for a solution. Spiking neural networks (SNNs) offer the \textit{potential} benefit of lower power consumption, on the condition that they are paired with appropriate hardware. Thus, in this paper we take a hardware-software co-design approach to create a neuromorphic audio denoising system that we call Zephyr. To do so, we take SOTA audio denoising algorithm Spiking FullSubNet (SFSN) \cite{hao2024spiking}, which we convert to a hardware friendly version via QAT and activation function simplification and batch norm removal, followed by mapping it to a custom digital hardware design - this is an accelerator circuit which could theoretically be taped out in silicon but for this work we use an FPGA to perform validation and gather latency estimates. However, SFSN, like many SOTA SNN algorithms, whilst benefiting from high activation sparsity, requires a range of different operation types to perform inference \cite{sun2024dpsnn} \cite{du2024ssmicassp}. Spiking layers can often make use of pure addition, but encoding layers can require matrix multiplication, and leakage terms often require elementwise multiplication. As such, in this work we also present a \textbf{heterogeneous} processing-element (PE) array, paired with an event-based memory read mechanism, to flexibly support the operations required by SFSN whilst also taking advantage of its activation sparsity.

\section{Background and Related Work}
\label{sec:Background}
Neural networks provide a data-driven way to improve on-device audio denoising algorithms that can be optimised against perceptual objectives. ANNs such as RNNoise \cite{valin2018rnnoise}, DCCRN \cite{hu2020dccrn} and FullSubNet \cite{hao2021fullsubnet} have shown that using a neural network in adjunction with traditional techniques can give an improvement in audio quality, however these ANNs require hardware support in the device upon which they are deployed. Specifically, large models require large memory resources and ANNs require a lot of dense MAC operations - this can make them awkward to deploy on highly power constrained SoCs like those found in wireless headphones. Recent algorithms such as SFSN \cite{hao2024spiking} and DPSNN \cite{sun2024dpsnn} seek to decrease computational complexity by using SNNs with sparse, binary activations. Whilst these have shown good theoretical power consumption and computational efficiency whilst maintaining high performance (15.20dB and 14.70dB on the Intel-DNS challenge dataset for SFSN and DPSNN respectively), SNNs require hardware that is able to support their operations whilst also being able to take advantage of their high activation sparsity and low computational complexity to be able to actualize these power consumption gains.

\begin{table*}[t!]
\centering
\caption{Speech-enhancement performance and energy efficiency
on the Intel N-DNS Challenge dataset \cite{timcheck2023ndns}. Energy costs for ANN/SNN baselines follow the Intel N-DNS operation-based energy-proxy
methodology.
}
\label{tab7}
\setlength{\tabcolsep}{4pt}
\begin{tabular*}{\textwidth}
{@{\extracolsep{\fill}}lccccc@{}}
\hline
\textbf{Model} &
\shortstack{\textbf{Year /}\\\textbf{Rank}} &
\shortstack{\textbf{Params (K)}\\\textbf{(precision)}} &
\shortstack{\textbf{SI-SNR}\\\textbf{(dB)} $\uparrow$} &
\shortstack{\textbf{DNSMOS}\\\textbf{OVR} $\uparrow$} &
\shortstack{\textbf{Energy}\\\textbf{Cost} $\downarrow$}
\\
\hline
\multicolumn{6}{l}{\textit{Official Intel N-DNS baselines}}
\\
Microsoft NsNet2~\cite{nsnet2}
& 2023 & 2,681 (FP32) & 11.63 & 2.95
& $12.51\,\mu\mathrm{J}$ \\
Intel DNS Network~\cite{timcheck2023ndns}
& 2023 & 1,901 (FP32) & 12.51 & 3.08
& --- \\
SDNN Network~\cite{timcheck2023ndns}
& 2023 & 525 (FP32) & 12.26 & 2.70
& $0.41\,\mu\mathrm{J}$ \\
\hline
\multicolumn{6}{l}{\textit{ANN baselines evaluated on Intel N-DNS}}
\\
Fast FullSubNet~\cite{fast_fullsubnet}
& 2023 & 1,141 (FP32) & 13.88 & 2.91
& $0.09\,\mathrm{mJ}$ \\
CMGAN~\cite{cmgan}
& 2024 & 1,413 (FP32) & 14.01 & 2.81
& $1.47\,\mathrm{mJ}$ \\

S5 (sparse-11)~\cite{pierro2025accelerating}
& 2025 & 500-4K(INT8) & 15.20 & ---
& $0.463\,\mu\mathrm{J}^{\dagger}$ \\
\hline
\multicolumn{6}{l}{\textit{SNN baselines evaluated on Intel N-DNS}}
\\

DPSNN~\cite{sun2024dpsnn}
& 2024 & 1400 (FP32) & 14.70 & 2.90
& $1.08\,\mu\mathrm{J}$ \\

\hline
\multicolumn{6}{l}{\textit{Intel N-DNS Challenge top-ranking systems}}\\
CTDNN LAVADL
& Rank 2 & 905 (FP32) & 13.52 & 2.97
& $1.76\,\mu\mathrm{J}$ \\
Sparsity SDNN
& Rank 3 & 344 (FP32) & 12.16 & 2.70
& $0.27\,\mu\mathrm{J}$ \\
PSNN
& Rank 4 & 724 (FP32) & 12.32 & 2.68
& $1.65\,\mu\mathrm{J}$ \\
Spiking-FullSubNet-XL
& Winner & 965 (FP32) & 15.20 & 3.03
& $1.48\,\mu\mathrm{J}$ \\
\hline
\textbf{Zephyr}
& \textbf{Proposed} & \textbf{965 (INT8)}
& \textbf{14.54} & \textbf{3.00}
& \textbf{52.9~nJ} \\
\hline
\multicolumn{6}{@{}p{\textwidth}@{}}{\footnotesize
\textsuperscript{$\dagger$}S5 sparse-11 achieves the same
15.20~dB SI-SNR as Spiking-FullSubNet-XL with $3.2\times$
lower effective compute and $5.37\times$ lower memory
less memory at iso-accuracy~\cite{pierro2025accelerating}. Its energy cost is estimated as $1.48/3.2=0.463\,\mu\mathrm{J}$ using the Intel N-DNS operation-based energy-proxy methodology, assuming that energy scales proportionally with effective operation count. This value is derived from the reported operation count rather than obtained from a direct hardware measurement.
The proxy is based on a 45-nm CMOS energy model of 4.6~pJ per FP32 MAC and 0.9~pJ per AC operation.}
\\
\end{tabular*}
\end{table*}

\section{Hardware-Friendly Denoising Algorithm}
\label{sec:pagestyle}

\subsection{Spiking-FullSubNet Architecture Overview}

\begin{figure}[h]
    \centering
    \includegraphics[
        width=\columnwidth,
        keepaspectratio
    ]{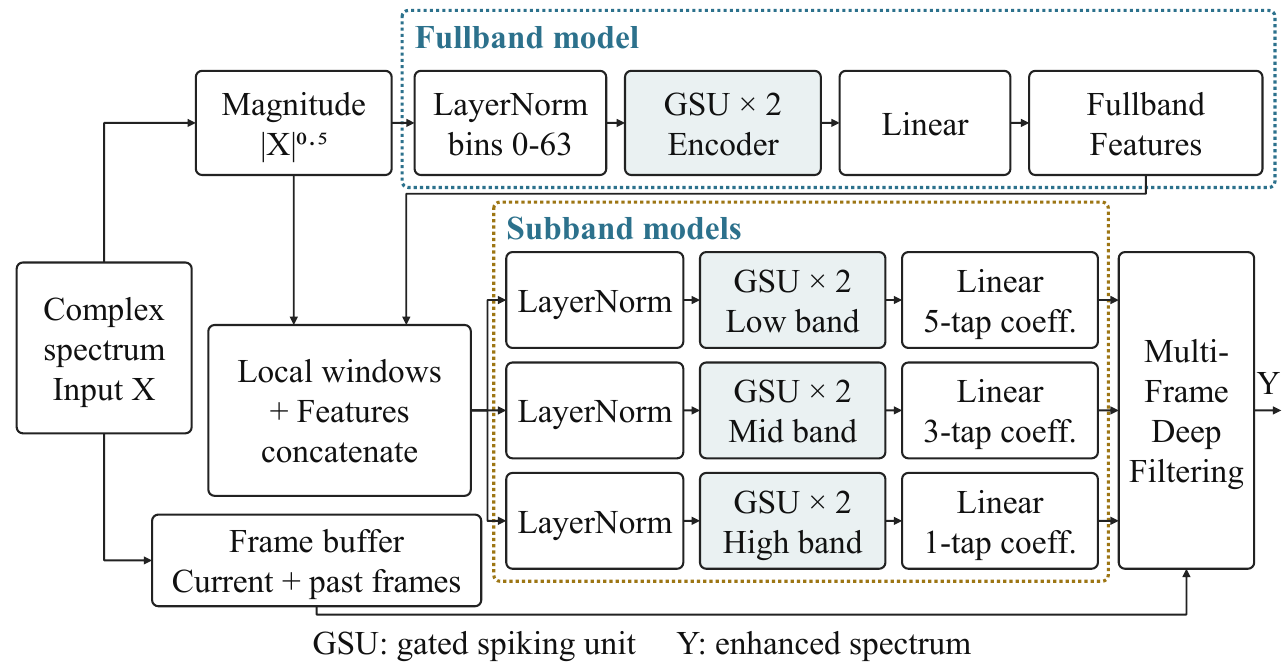}
    \caption{Model data flow of Spiking-FullSubNet. }
    \label{fig:zephyr-model}
\end{figure}

In this work, we build on SFSN, whose building-block element is
the gated-spiking neuron (GSN), as described in
Sec.~\ref{sec:gsn}. SFSN combines a full-band GSN encoder with
three sub-band branches, as seen in Fig.~\ref{fig:zephyr-model}. 

The full-band branch extracts a global
spectral embedding, while each sub-band branch combines this
embedding with neighboring frequency bins to capture local
spectral structure. The low-, mid-, and high-band branches
predict complex filtering coefficients with 5, 3, and 1 taps,
respectively. Multi-Frame Deep Filtering (MFDF) applies these
coefficients to the current and preceding STFT frames to
reconstruct the enhanced spectrum.
This data flow requires dense matrix--vector products,
spike-driven accumulation, gated state updates, batch and layer normalization,
and complex multiply--accumulate operations.

\subsection{GSN Membrane Update: Original Formulation}
\label{sec:gsn}

The original GSN neuron equation, before quantization, is
given by
\begin{align}
i^l(t) &= W_{mn}^l o^{l-1}(t) + W_{nn}^l o^l(t-1) + b^l, \label{eq:input_current}\\
\lambda^l(t) &= \sigma\big(W_{mn}^l o^{l-1}(t) + W_{nn}^l o^l(t-1) + \tilde{b}^l\big), \label{eq:gate}\\
u^l(t) &= \lambda^l(t)\, u^l(t-1) + \big(1-\lambda^l(t)\big)\, i^l(t). \label{eq:membrane_update}
\end{align}
where $\lambda^l(t)$ is a dynamic decay gate,
$u^l(t-1)$ is the previous membrane potential,
$i^l(t)$ is the input current. The gate controls the relative
contributions of the previous state and the current input.
In the floating-point implementation, each membrane update
involves two element-wise multiplications. Zephyr preserves
this update structure while implementing the gating products
and subsequent rescaling with integer arithmetic.

\subsection{Hardware-Friendly Implementation Methods}
\label{sec:membrane-schemes}

For edge deployment, Zephyr applies QAT-based symmetric signed
INT8 quantization to the neural-network weights, all non-spiking
activations, and recurrent membrane states. Matrix operations
accumulate in INT32. Linear and GSN matrix weights use
per-output-channel scales, whereas continuous
activations use per-tensor scales and recurrent membrane states
use per-channel scales. Binary spikes are represented separately
from these continuous activations. The quantization step reduces weights from 4MB to 1MB, this is an important step as 4MB would require an unfeasible amount of SRAM for a low power edge SoC like those used in hearing aids or bluetooth headphones.
For the GSN membrane update, the two gated products use wider
integer intermediate precision. A 255-entry sigmoid LUT maps
signed preactivations in $[-127,127]$ to unsigned 8-bit gate
coefficients with denominator 255, avoiding runtime exponential
evaluation. Precomputed integer multipliers and shifts align the
scales of the gated products and subsequent state updates. The
batch-normalization affine transform following the membrane
update is merged with the corresponding bias and
requantization parameters to further decrease computation. This folding is distinct from the
data-dependent LayerNorm operations in the full-band and
sub-band paths.

\subsection{Energy Analysis Methodology}
\label{sec:energy-analysis}

For comparison with prior Intel N-DNS Challenge submissions,
the operation-based power proxy is expressed
as~\cite{timcheck2023ndns,hao2024spiking}
\begin{equation}
P_{\mathrm{proxy}}
= \mathrm{SynOps} + 10 \times \mathrm{NeuronOps},
\label{eq:proxy}
\end{equation}
where $\mathrm{SynOps}$ and $\mathrm{NeuronOps}$ denote
synaptic and neuron operation rates, respectively.
The weighting factor represents the relative cost assigned
to neuron operations in the proxy. This metric captures
changes in operation counts but does not explicitly account
for operand precision, LUT access, or channel-wise
requantization. Computational energy is estimated by weighting operation
counts after zero-skipping by their corresponding costs.
For 45\,nm, 0.9\,V CMOS, INT8 multiplication and INT32
addition are assigned 0.20\,pJ and 0.10\,pJ per operation,
respectively~\cite{horowitz2014computing}.
Full-model accounting must also include membrane updates,
channel-wise rescaling, LUT access, normalization, and
deep filtering; these arithmetic coefficients alone do
not capture memory-access costs.
Since SFSN already accounts for spike
activity~\cite{hao2024spiking}, consistent counting rules
are required to avoid double-counting sparsity savings.
This cost-model estimate does not represent measured
FPGA energy, rather the computation energy that could be achieved in custom silicon.

\section{Sparsity-Aware Flexible Heterogenous FPGA Accelerator}
\label{sec:typestyle}
\subsection{Accelerator Architecture}
\begin{figure}[t]
    \centering
    \centerline{\includegraphics[width=8.5cm]{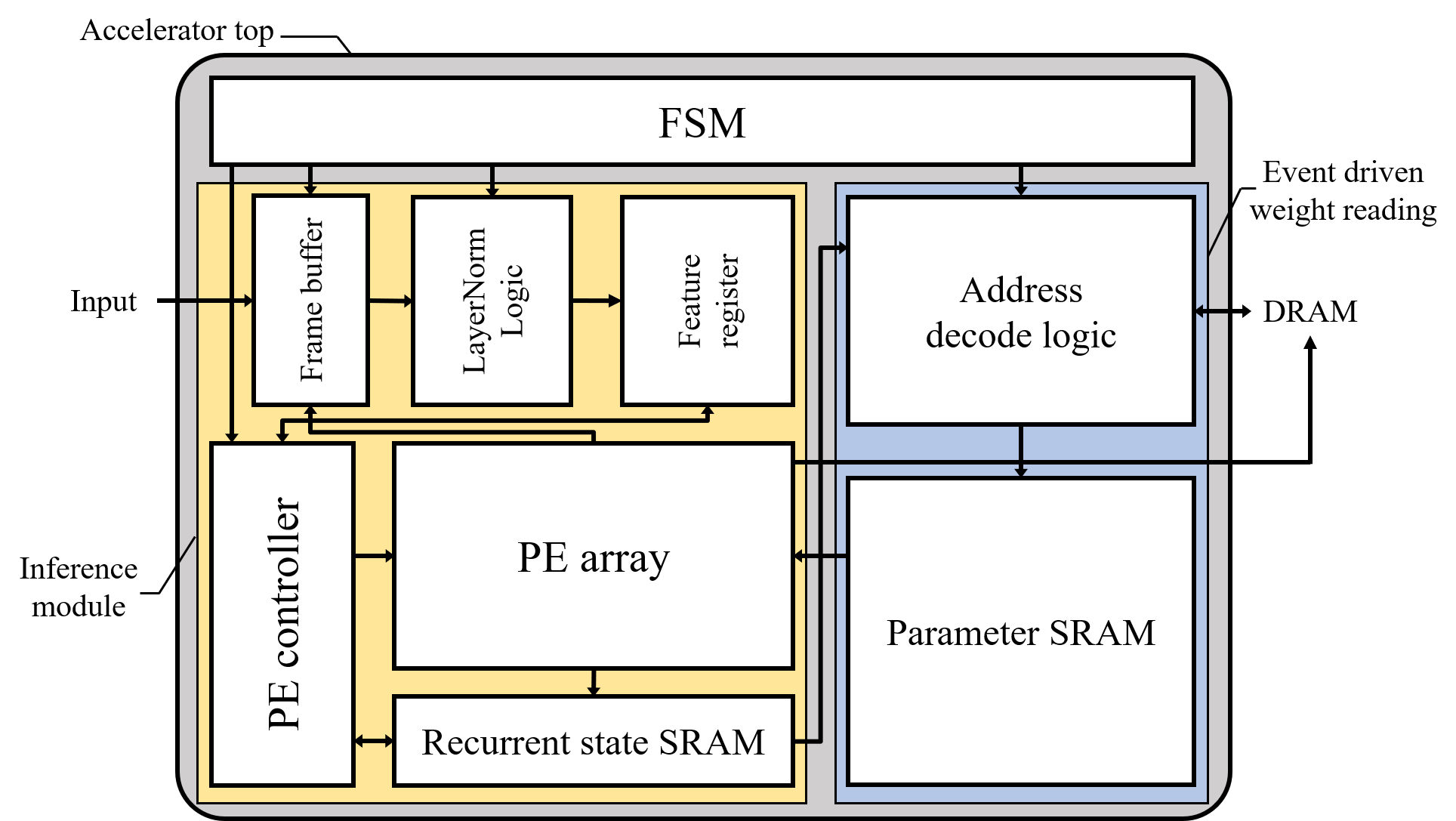}}
    \caption{Overall architecture of the Zephyr accelerator.}
    \label{fig:Accelerator block diagram}
\end{figure}

As illustrated in Fig. \ref{fig:Accelerator block diagram}, the proposed Zephyr accelerator comprises three parts: an inference module, a finite state machine and a event driven weight reading memory. A top-level finite-state machine (FSM) coordinates their operation. The inference module executes the hardware-friendly audio denoising algorithm described in Section 3. It integrates frame buffering, layer normalization, feature register, a processing-element (PE) array and recurrent-state SRAM. The PE controller schedules computations on the PE array across the different inference stages. The event driven weight reading memory subsystem manages weight storage and access. It comprises a address decoding logic, and BRAM-based parameter memory. During matrix–spike computation, the address decoding logic translates stored active-neuron indices into the corresponding DRAM addresses for weight retrieval. The parameter memory buffers on-chip weight storage and supplies the PE array with the weights required for computation. By restricting weight retrieval and accumulation to active neuron events, this subsystem exploits activation sparsity to avoid unnecessary DRAM accesses and PE operations.

\subsection{PE Array Architecture}
\begin{figure}[t]
    \centering
    \includegraphics[width=\linewidth]{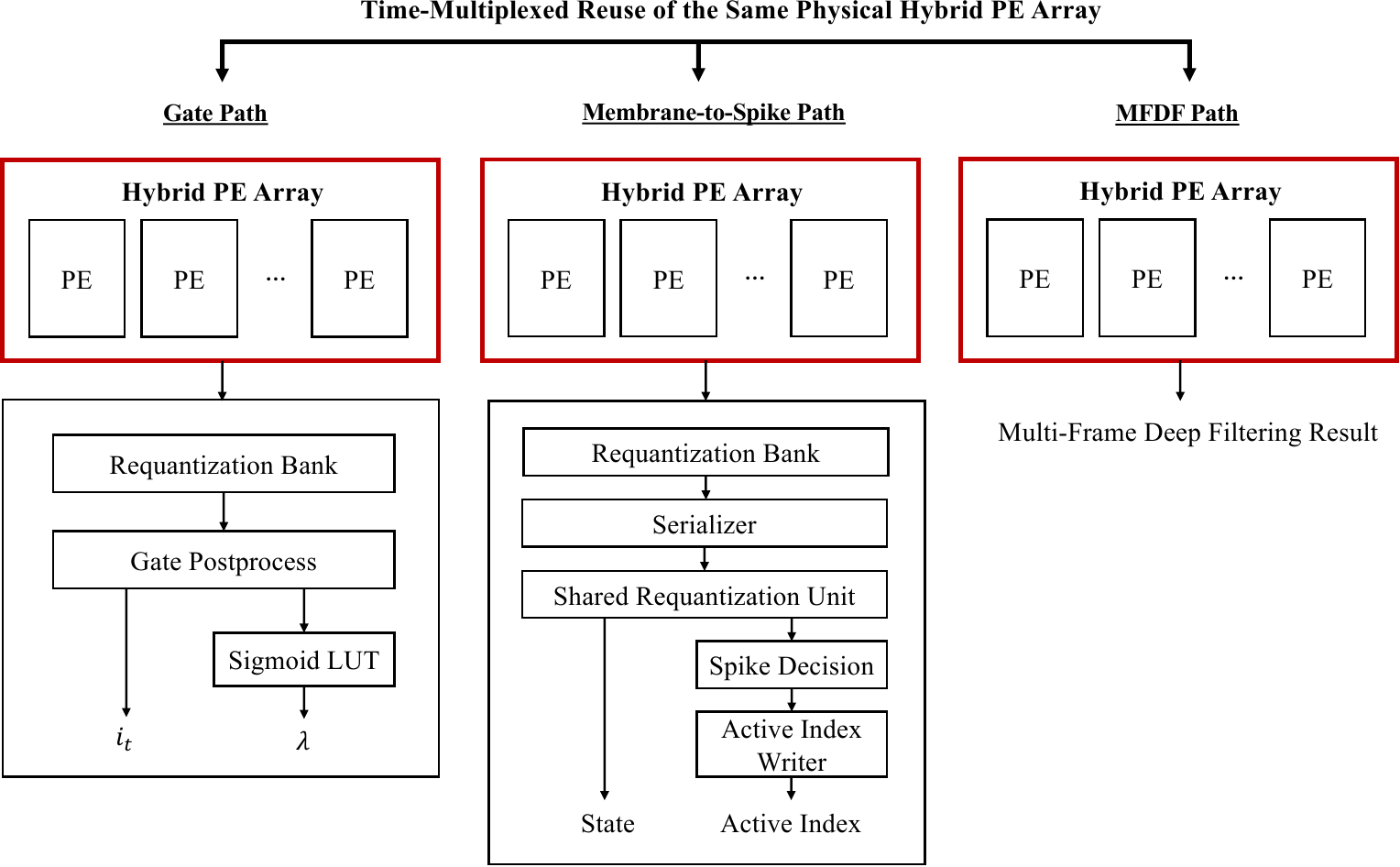}
    \caption{Time-multiplexed reuse of the same physical Hybrid PE array across different computation paths.}
    \label{fig:pe_reuse}
\end{figure}

Unlike conventional neural network accelerators that mainly target homogeneous multiply-accumulate (MAC) operations, this SNN, like many others from the literature, has a highly heterogeneous computation pattern. Thus, we design a flexible PE array capable of supporting multiple types of operations, including matrix multiplication, pure addition and element-wise multiplication (Multi-Frame Deep Filtering). As illustrated in Fig.~\ref{fig:pe_reuse}, the same Hybrid PE array is reused via time-multiplexing across different computation phases. The three PE arrays shown in the figure represent different usages of the same physical hardware, rather than three independent PE arrays. During gate computation, the PE array performs the required MAC operations, after which the accumulated results are processed by the requantization bank and gate postprocessing unit. The gate value is then mapped through the sigmoid LUT to generate the corresponding gating coefficient. For membrane and spike firing, the same PE array is reused to compute the membrane-state update. The results are then serialized and processed by the shared requantization unit before spike generation. The spike decision unit identifies active neurons, and their indices are recorded by the active-index writer for subsequent sparse matrix--spike computation. For MFDF, the Hybrid PE array is reused to perform complex MAC accumulation. Real and imaginary components are mapped onto pairs of PEs and accumulated over multiple filter taps, allowing multi-frame deep filtering to be implemented without introducing a dedicated complex arithmetic module. By time-multiplexing the Hybrid PE array across these heterogeneous computation phases, the proposed architecture avoids duplicating arithmetic resources for operation-specific modules. This reuse improves hardware utilization while keeping the flexibility required by the different computation patterns in this neural network.

\begin{figure}[t]
    \centering
    \begin{minipage}[t]{0.48\linewidth}
        \centering
        \includegraphics[
            width=\linewidth,
            height=3.6cm,
            keepaspectratio
        ]{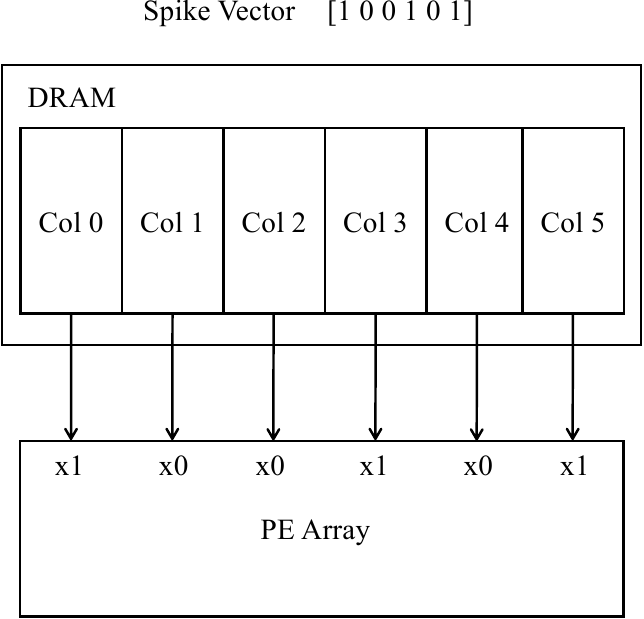}

        \vspace{1mm}
        {\footnotesize (a) Dense execution}
    \end{minipage}%
    \hfill
    \begin{minipage}[t]{0.48\linewidth}
        \centering
        \includegraphics[
            width=\linewidth,
            height=5cm,
            keepaspectratio
        ]{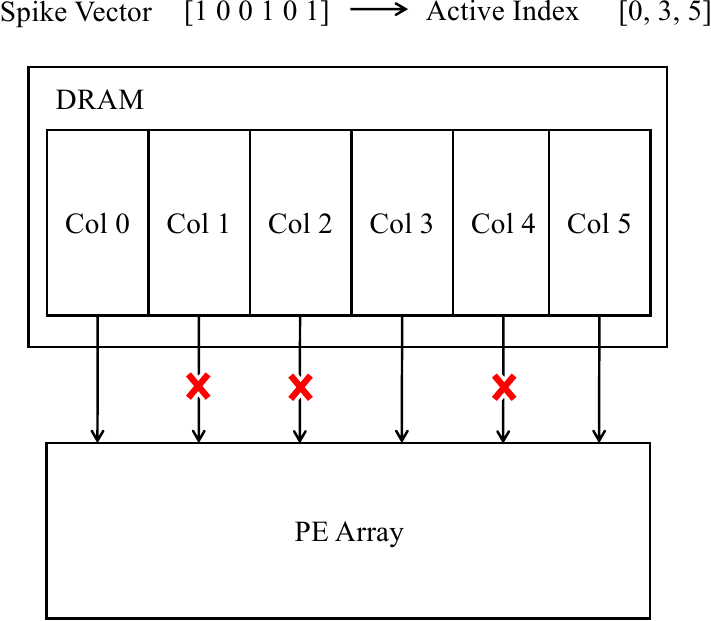}

        \vspace{1mm}
        {\footnotesize (b) Sparsity-aware execution}
    \end{minipage}

    \caption{Comparison between dense matrix--spike execution and the proposed sparsity-aware execution. In the proposed design, only weight vectors corresponding to active spike indices are fetched and accumulated, eliminating DRAM accesses and PE operations associated with inactive spikes.}
    \label{fig:sparsity_comparison}
\end{figure}

\subsection{Sparsity-Aware Computation}
Our spiking neural network exhibits high activation sparsity. To exploit this property, we introduce a sparsity-aware mechanism that skips inactive neurons during matrix--spike multiplication. Rather than processing all spike entries, only the indices of active spikes are recorded and scheduled for computation, allowing the PE array to fetch and accumulate weights exclusively for active spike events. As a result, computations and DRAM accesses associated with inactive spikes are eliminated. As illustrated in Fig.~\ref{fig:sparsity_comparison}, dense execution process all spike entries regardless of its activity. Consequently, all weight vectors are fetched and sent to PE array even when the corresponding spike value is 0. In contrast, the proposed sparsity-aware mechanism first extracts the indices of all active spike. Only the weight vectors corresponding to active index are fetched and scheduled for computation. Since inactive spike entries are removed before weight fetching, both unnecessary weight access and PE operations are avoided.

\section{Experimental Setup}
\label{sec:majhead}

The accelerator was implemented on a PYNQ-Z1 FPGA with
32 PEs operating at 100 MHz and evaluated using the Intel
N-DNS Challenge dataset. Speech-enhancement quality was
evaluated using SI-SDR, while hardware performance was
measured using board-loop latency and sustained real-time factor.
Table~\ref{tab:fpga_results_comparison} summarizes the implementation results of
the proposed accelerator as well as comparison against prior works.

\begin{table}[t]
\centering
\caption{FPGA implementation summary and comparison with speech enhancement FPGA accelerators.}
\label{tab:fpga_results_comparison}

\footnotesize
\renewcommand{\arraystretch}{1.12}
\setlength{\tabcolsep}{2.5pt}

\begin{tabular*}{\columnwidth}
{@{\extracolsep{\fill}}lclc@{}}
\toprule
\multicolumn{4}{c}{\textbf{Zephyr FPGA Implementation Summary}} \\
\midrule

\textbf{Metric} & \textbf{Value} &
\textbf{Metric} & \textbf{Value} \\
\midrule

Platform
& PYNQ-Z1
& Device
& XC7Z020-1 \\

Clock frequency
& 100 MHz
& Number of PEs
& 32 \\

Mean latency
& 5.382 ms
& Request/op. skip$^{a}$
& 87.85\% \\

Traffic reduction$^{b}$
& 54.47\%
& &
\\

\end{tabular*}

\begin{tabular}{l c r r r c c}
\toprule
\multicolumn{7}{c}{\textbf{Comparison with Speech Enhancement FPGA Accelerators}} \\
\midrule

\textbf{Work}
& \textbf{FPGA}
& \textbf{LUT}
& \textbf{BRAM$^{\dagger}$}
& \textbf{DSP}
& \textbf{RTF}
& \textbf{SNN} \\
\midrule

Flamis et al.~\cite{flamis2022fpga}
& XC7Z020
& 30,074
& 117.5
& 194
& --
& \ding{55} \\

Abdullah et al.~\cite{abdullah2024hardware}
& XC7A100T
& 33,910
& 35.6
& 0
& --
& \ding{55} \\

Ni et al.~\cite{ni2026cstrn}
& XCZU48DR
& 89,987
& 903
& 83
& 0.579
& \ding{55} \\

\textbf{Zephyr}
& XC7Z020
& \textbf{44,717}
& \textbf{78.5}
& \textbf{62}
& \textbf{0.727}
& \ding{51} \\

\bottomrule
\end{tabular}

\vspace{1mm}

\begin{minipage}{\columnwidth}
\footnotesize
$^{a}$Skip zero-valued operations.
$^{b}$Reduction in parameter data read from memory.
$^{\dagger}$BRAM usage is normalized to 36-Kb BRAM equivalents.
\end{minipage}

\end{table}
 
\section{Conclusion}
\label{sec:print}
In this work we demonstrate with our system Zephyr that a neuromorphic hardware-software co-designed audio-denoising system is capable of performing high quality audio denoising whilst achieving low power and processing latency.




\section{Acknowledgments}
This work was supported by the National Science and
Technology Council, Taiwan, under contract no. NSTC 115-2223-E-007-010 and NSTC 115-2640-E-007-014.

\bibliographystyle{IEEEbib}
\bibliography{references}
\end{document}